\documentclass[10pt,letterpaper]{article}
\usepackage{geometry}
\usepackage{amsmath}

\usepackage[explicit]{titlesec}

\titleformat{\section}
{\normalfont\normalsize\bfseries\centering}{\thesection}{1em}{\MakeUppercase{#1}}

\usepackage[utf8]{inputenc}

\usepackage{cite}

\usepackage{nameref,hyperref}

\usepackage{microtype}
\DisableLigatures[f]{encoding = *, family = * }

\usepackage{changepage}

\usepackage[aboveskip=1pt,labelfont=bf,labelsep=period,singlelinecheck=off]{caption}

\makeatletter
\renewcommand{\@biblabel}[1]{\quad#1.}
\makeatother

\usepackage{lastpage,fancyhdr,graphicx}
\usepackage{epstopdf}

\usepackage{color}

\definecolor{Gray}{gray}{.25}

\usepackage{graphicx}

\begin{document}
\vspace*{0.35in}

\begin{center}
{\Large \textbf \newline{Magnetar Crusts: Matter at $10^{15}$ Gauss}}
\newline
\\
Vikram Soni\textsuperscript{1,2,*},
Mitja Rosina\textsuperscript{3}

\bigskip
\bf{1} Centre for Theoretical Physics, Jamia Millia Islamia, New Delhi 110025, India
\\
\bf{2} Inter University Centre for Astronomy and Astrophysics, Post Bag 4, Pune 411007, India
\\
\bf{3} Faculty of Mathematics and Physics, University of Ljubljana, Jadranska 19, SI-1000,\\ 
Ljubljana, Slovenia\\
and Jo\u zef Stefan Institute, Ljubljana, Slovenia
\\
\bigskip
*vsoni.physics@gmail.com
\end{center}

\section*{Abstract}

In this work we find that the at the  high polar magnetic fields of magnetars, $B \sim 10^{14-15}$ G, the outermost crust ($\rho < 10^7$ gm/cc) of the star can become a transverse insulator and a filamentary crystal along the field direction. Also, the transverse conductivity in the crust goes  inversely as the square of  polar magnetic field (as $1/B^2$). At these high fields the transverse crustal currents associated with the polar magnetic field can then dissipate more effectively via Ohm's law.  
\newline

\section{Introduction}

Till recently, the majority of neutron stars were considered to be pulsars which carry an inherited or fossil polar magnetic field, of the order of $B \sim 10^{9-12}$ G, from their erstwhile collapse  . However, now we have an increasing population of magnetars which are usually isolated neutron stars with the largest known polar magnetic fields ($10^{14} - 10^{15}$ G) that have much smaller spin down ages of $10^{3}$ - $10^{5}$ years. Unlike pulsars, they are distinguished by the emission of a quiescent radiative X-ray luminosity of $10^{34}$ - $10^{36}$ erg/s. Besides, some of them emit repeated flares or bursts of energy typically of $10^{42} - 10^{44}$ erg, and at times of even higher intensity \cite{Hurley,Palmer}. The periods of magnetars fall in a surprisingly narrow window of 2-12 s. \emph{However, there are exceptional magnetars which may not share all these features.}\\

At such large periods, the energy emitted in both quiescent emission and flares far exceeds the loss in their rotational energy through dipole radiation. The most likely energy source for these emissions is their magnetic energy, yet there is no evidence of a decrease in their surface (polar) magnetic fields with time \cite{Thompson}. There have been many attempts to explain some of this physics of which the most canonical is the magnetar model of Duncan and Thompson \cite{DuncanThompson,ThompsonDuncan}, which is known as the {\em dynamo mechanism for magnetars}. This model requires the collapse of a large mass progenitor to a star which starts its life with a period close to a millisecond. At the high temperatures generated by the collapse process, such a fast rotation can amplify the inherited pulsar valued field of $10^{12}$ G to $10^{15}$ G or more. However, as described below, several observations on magnetars are hard to understand from such a  model.\\

If magnetars are born with their high surface magnetic fields, then after a flare, one would expect a decrease in the magnetic energy and consequently a {\em decrease} in the polar magnetic field of the magnetar, accompanied by a fall in the dipole radiation mediated spin down rate. However,this is not the case and sometimes the opposite is seen: the spin down rate {\em increases} after the flare indicating an increase in the surface magnetic field \cite{DarDR,Dar,Marsden,Kaspi}. Further, inspite of the magnetic energy loss from steady X-ray emission, the magnetic field of magnetars appears to remain high all the way till the end of spin down, when the stars have the largest periods.\\

In an earlier work \cite{DipSoni,mag}, it was shown that it may be  possible to explain many unusual features of magnetars if they have a core with a large magnetic moment density, created by a strong interaction phase transition in the high density core at birth. Initially, the core magnetic field is shielded by screening currents set up by the change of flux in the electron plasma in and around the core. In time, the screening currents dissipate till finally the field emerges at the surface of the star. In this model, the field increases with time till it reaches its relaxed state after the screening currents have dissipated.\\

Whereas, it is the dynamo currents that dissipate in the first model, in the screened core model \cite{DipSoni,mag,Soni3}, above, it is the screening currents that dissipate. The effect on the polar magnetic field is the opposite. In the dynamo model, $B_{polar}$ goes down through dissipation with time. In the screened core model, $B_{polar}$ goes up as the screening currents dissipate.\\


We note that at the high polar crustal and surface magnetic fields, $B_{polar} > 10^{13 - 14}G$, the Landau radius becomes smaller than the Bohr radius.
When the magnetic fields go up $10^{15(14)}$ G), the electrons in the outer shell of the crust (  for density $ \sim 10^7 $gm/cc) are localised, approximately in a landau radius which is much smaller than the inter ionic spacing.  We  then have get deformed cigar like ions which are almost neutral. 
The coulomb repulsion between the ions comes down  making the crystal binding weaker. The screening length becomes smaller than the inter atomic distance \cite{ShaRed,Bed} in the transverse direction. The overlap between the electron wavefunctions between sites in the transverse direction diminishes with a simultaneous drop in the transverse conductivity.  We find that at such high magnetic fields the outermost crust ($\rho < 10^7$ gm/cc) of the star can become a transverse insulator and a filamentary crystal along the field direction\\

Further, we find that the at the typical high polar magnetic fields of magnetar, $ B \sim 10^{14-15} $ Gauss, the transverse conductivity in the crust goes  inversely as the square of  polar magnetic field (as $1/B^2 $). At these high fields, this indicates that the transverse crustal currents associated with the polar magnetic field can then dissipate more effectively via Ohm's law.  This could explain the anomalously high X-ray luminosity of magnetars.\\


\section{The ground state at high polar magnetic fields ($ B \sim 10^{14-15} $ G): Landau levels}

When the external magnetic field rises to above $10^{13}$ -- $10^{14}$ G, the magnetic field dominates the motion in the direction transverse to the field as the atoms transform to a cylindrical shape along the field. This happens when  the Landau radius in the lowest Landau level, $ r_L = \sqrt{hc/(2 \pi B e)}$
becomes smaller than the Bohr radius of the most tightly bound electron orbit, $ a_0/Z = \hbar^2/(e^2 m_e) \times (1/Z)$. However, the field starts influencing the electron dynamics even at $B \sim 10^{12}$ G.\\

In the presence of a constant magnetic field, the electron states in the plane transverse to the magnetic fields are called Landau levels. The lowest Landau level has a large degeneracy. This degeneracy is given by  $ABe/(hc)$ where $A$ is the area over which magnetic field is deployed, and $\Phi_0 = hc/e $ is the unit quantum flux.\\

First we need to write down the electron density, $n_e $, when only the lowest Landau level is fully occupied in the plane that is transverse to the magnetic field and the ground state is a one dimensional Fermi sea along the direction of the field,

\begin{equation}
n_e =  n_{t{2d}} \times n_{1d}
\end{equation}
where $n_{1d}  = k^f/(2\pi)$ is the 1 dimensional electron Fermi sea density along the direction of the magnetic field and $n_{t{2d}} =  Be/(hc)$ is the 2 dimensional transverse electron density or the number of electrons per unit area. Also, $k^f = n_e h c\cdot 2\pi/( B e)$ and the Fermi momentum, $p_{z}^f = (h/2\pi) k^f $.\\

Now the expression for the relativistic  electron energy eigenvalues in an external magnetic field is given in Shapiro and Dong Lai \cite{shapiro} and Dong Lai \cite{Dong Lai},

\begin{equation}
E_{\nu,p_z} = \sqrt{(p_z c)^2 + m^2c^4 (1 + \nu \frac{2B}{B_c})}
\end{equation}

where $m$ is the electron mass, B is the external magnetic field, $\nu$ is the order of the Landau level that takes integer values starting from $\nu = 0 $, and $B_c  =  m^2 c^3 /(h/(2\pi) \times e)$.\\

The Fermi energy for the zeroth Landau level is given by,

\begin{equation}
E_f ^0  = \sqrt{(p_z^f c)^2 + m^2 c^4}
\end{equation}

where $p_z^f c = h k^f/(2\pi) = n_e h^2 c/( B e)$ and the energy gap between the first and the zeroth Landau level is,

\begin{equation}
\Delta  = \sqrt{ (p_z^f c)^2 + m^2c^4 (1  + \frac{2B h e}{2\pi m^2c^3})} - \sqrt{(p_z^f c)^2 + m^2c^4 }
\end{equation}

The condition for the sole occupation of the lowest Landau level is then

\begin{equation*}
\frac{E_f^0}{\Delta} < 1
\end{equation*}

\begin{center}
or
\end{center}

\begin{equation*}
\sqrt{ 1 + \frac{2B c h e}{2\pi (1 + (p_z^f c)^2/ ( m^2 c^4))}} - 1 >1
\end{equation*}

After a little algebra, this condition maybe written as

\begin{equation}
2 \leq \sqrt{1 + \frac{2B h e}{m^2 c^3 2\pi (1 +\kappa)}}
\end{equation}

where $\kappa = n_e^2 h^4/(B^2 e^2 m^2)$.\\

For  $  B =  10^{15}$ G, this yields,

\begin{equation}
n_e \leq 3.8 \times 10^{31}~\text{/cc}
\end{equation}

This translates to an average mass density,

\begin{equation}
\rho \leq  6.3 \times 10^7~\text{g/cc}
\end{equation}

At this density, we find that the Fermi energy, $E_f $ is 1.97 MeV, which falls in the relativistic regime. This further implies that for fields $B = 10^{15}$ G, we are in the nucleon capture regime between $^{62}$Ni and $^{64}$Ni nuclei  in the crust.\\

From the same set of steps for fields $B = 10^{14}$ G, we find the condition for the sole occupation of the lowest Landau level is,

\begin{equation}
n_e  \leq  0.7  \times 10^{30}~\text{/cc}
\end{equation}

which translates to a mass density,

\begin{equation}
\rho \leq 1.15 \times 10^6~\text{g/cc}
\end{equation}

We note that the value for the kinetic energy, $p_{z}^f c \sim  0.36$ MeV which is somewhat less than the mass energy. This establishes that here we are not in a relativistic regime. This falls in the outer crust where, $^{56}$Fe is the favoured nucleus.\\

We have found that if we have the condition that only the lowest (zeroth) Landau level is occupied, we have an upper limit for the nucleon density which is compatible with only the outer shell of the crust of the neutron star, $\rho  \sim 4 \times 10^{7} - 1.2 \times 10^{6}$ gm/cc. At higher densities the next Landau level gets occupied. We shall discuss the significance of this shortly.
\newline

\subsection{The Crystal}

The atoms in the crust are strongly compressed by gravity and get completely ionised so that the electrons form a degenerate Fermi sea at magnetic fields characteristic of pulsars $\sim 10^{12}$ G. In the outer crust, approximately a little less than half the number (Z) of nucleons in the ions are protons (charge neutrality requires the number of protons must be the same as the number of conducting electrons) and the rest are neutrons. This results in a strong Coulomb repulsion between the ions. It is the interplay between gravity and Coulomb repulsion that makes the crystalline crust. The nucleon density is a little more than the twice the electron density. Starting with $^{56}$Fe ($10^6$ g/cc) at the outer crust, as density increases inward, we go through the neutron capture regime to $^{64}$Ni and then all the way to neutron rich $^{118}$Kr ($10^{11}$ g/cc) at the inner crust (see \cite{Baym} and references therein).\\

Now let us examine the inter-ionic distance in the crust and compare it with the Landau radius for the threshold density for occupation of the lowest Landau level for magnetic fields.\\

For external fields of $B= 10^{15}$ G, the electron density at the threshold of occupation of only the lowest Landau level is $n_e \leq 3.8 \times 10^{31}$ /cc. Given that this neutron capture regime is made up of $^{63}$Ni nuclei, we have a nucleon density of $2.25$ times the electron density and an inter-ionic distance of $\sim 900$ fm whereas the Landau radius, which is independent of the electron density as it depends only on B, is $r_l \sim 80$ fm. We thus expect that all the electrons can be attracted to a smaller cylinder around the positive ions that are in a one dimensional array along the magnetic field lines. The size of electron wave functions is of the same order as the Landau radius \cite{Dong Lai}.\\

For external fields of $B= 10^{14}$ G, the electron density at the threshold of occupation of only the lowest Landau level is $n_e \leq 0.7 \times 10^{30}$ /cc. Given that this crustal regime is made up of $^{56}$Fe nuclei, we have a nucleon density of $2.15$ times the electron density and an inter-ionic distance of $\sim 3300$ fm whereas the the Bohr radius is only $\sim 300$ fm. We thus expect that all the electrons can be attracted to a smaller cylinder around the positive ions along the magnetic field lines.\\

This implies that for those densities where only the lowest Landau level is occupied we are in a regime where the Landau radius is about ten times smaller than the inter-ionic distance. Though the electrons in the Landau level do not have a fixed centre or axis, in this case, they will crowd about the ions as they are attracted to ions and can thus reduce the Coulomb energy of the ions. One effect of this is that the original crystal is lost and it will recast itself into a one dimensional crystal along the field but lose its moorings in the transverse direction and it will resemble strands of spaghetti and also turn from a conductor to an insulator. This large distance between the strands in the transverse direction means that the neutral cylindrical atoms do not feel much Coulomb repulsion between them in the transverse plane. They can then flop around or oscillate and will feel the Coulomb repulsion if they approach their transverse neighbours. In this regime, the system is termed strongly quantizing (Landau). Since the electrons are tightly bound along the magnetic field lines, the transverse conductivity is very low.\\

At densities higher than the threshold density, further inside the crust, the next Landau level will get occupied, marginally restoring some conductivity. Even at higher density when the second Landau level is occupied, the system continues to be Landau quantizing - in other words the magnetic field strongly influences the dynamics. The wavefunction spreads out somewhat but the conductivity stays low. However, at even higher density, when many Landau levels are occupied, the system will become non quantizing and the effect of the magnetic field diminishes.The wavefunctions spread out, mimicking a Fermi sea and restoring conductivity. In fact a detailed account of this phenomenon is given in Dong Lai's paper \cite {Dong Lai}. We now go on to look at the behaviour of the transverse conductivity at high polar field which determines the ohmic dissipation in the crust. This is responsible for the a large part of the X-ray luminosity of the magnetars.
\newline

\subsection{Transverse Conductivity Dependence on Magnetic Field}

In this section we will briefly review the electrical  conductivity at high magnetic fields. Whatever be the model for magnetars, we know that it is the dissipation of the magnetic field energy that fuels the X-ray luminosity, which is a distinguishing feature of magnetars. Further, the polar magnetic field depends on currents that run in the plane transverse to the field. Hence it is the transverse conductivity that will determine the dissipation of these currents, which in turn influence the polar field. In an  early work,  Haensel et al \cite{Haensel}, find that the electrical conductivity has an inverse dependence on the ambient magnetic field. We point to some further work on this subject\\

The main charge transport in the crust is due to electron currents. In what follows, we write the equations governing electron motion in a medium with electric and magnetic fields. These results are for the single particle equations where the direct effects of the Fermi sea have not been factored in, but are replaced by an average equilibrium velocity called the drift velocity that describes the system. Such a situation is often likened to the Druid conductivity. For high polar magnetic fields in the magnetar ball park, $B_p > 10^{14}$ G, the isotropic Fermi sea gets deformed into Landau levels in the direction transverse to the magnetic fields but continues as a one dimensional Fermi sea along the direction of the magnetic field. Above, we have considered the threshold densities below which only the lowest Landau level is occupied  for magnetar strength magnetic fields. Due to the large degeneracy of the lowest Landau level, all the electrons in this level are in the same state. Thus, a single particle model for electrons in a magnetic and electric field can be used (Druid model) to describe the dynamics \cite{Soni3}. We have

\begin{equation}
m^* \Big(\frac{d\vec v}{dt} +\frac{ \vec v}{\tau}\Big)  =  -e\vec E + \frac{\vec v}{c} x \vec B
\end{equation}

where $\vec v$ is the drift velocity, $ \tau$ is the relaxation time (or collision time), and $m^*$ is the effective mass.\\

In the steady state, $\frac{d\vec v}{dt} = 0$.\\



Defining, $\sigma_0 = n e \lambda$, as the isotropic conductivity in the absence of the magnetic field \\
where, $n$ is the electron density and $e$ is the electric charge  and 
$\lambda = e \tau/ {m^*}$  




we can use the following the transverse conductivities as given in \cite{Soni3}, with, $\alpha = (1 + \beta^2 )$ and $ \beta = \lambda B_z/c$ \\

\begin{align}
\sigma_{xx} & =\sigma_{yy} = \frac{\sigma_0}{\alpha}\\
\sigma_{xy} & = -\sigma_{yx} = -\frac{\sigma_0 \lambda B_z}{c \alpha}\\
\end{align}

It is to be noted that the transverse conductivities depend on the magnetic field $B_z$, whereas the isotropic conductivity does not. We can now write down the Ohmic magnetic field decay times,

\begin{align}
{\tau^D}_{ohm} &= \frac{ 4\pi\sigma_0\cdot {L^2}}{ c^2 \alpha} \sim \frac{4\pi(n^2 e^2)\cdot {L^2}}{\sigma_0 B^2}~\text{and}\\
\sigma_{xx} &=\sigma_{yy} = \frac{\sigma_0}{ \alpha}
\end{align}

where ${\tau^D}_{ohm}$, is the diagonal transverse ohmic dissipation time and we have used $\lambda =\sigma_0/(n e) $ 

Similarly

\begin{align}
\tau^{ND}_{ohm} &= \frac{ 4\pi\sigma_0 L^2}{c^2 \alpha} \times\beta~\text{and}\\
\sigma_{xy} &= -\sigma_{yx} = -\frac{\sigma_0 \lambda B_z}{c \alpha}
\end{align}

where ${\tau^{ND}}_{ohm}$ is the non diagonal( Hall) transverse dissipation time.\\

From the above, we find that the conductivity tensor in this Landau quantized regime is highly anisotropic. The parameter that controls the conductivities is $\beta = \frac{\lambda B_z}{c}$. If $\beta \gg 1$, then we can neglect the factor of unity in, $\alpha = (1 + \beta^2)$ and the conductivity becomes magnetic field dependent. For large fields, it is the second term in $\alpha = (1 + \beta^2 )$ that dominates, eg., $\sigma_{xx} =\sigma_{yy} = \sigma_0/\alpha$, go inversely as the \textsl{square} of the magnetic field ($1/B^2$). On the other hand, the non diagonal conductivity, $\sigma_{xy} = -\sigma_{yx}$ goes as ($1/B$).\\

 There is also work  that considers  the conductivity at typical magnetic fields in the range $B \sim 10^{12-14}$ G, in the non-Landau quantizing regime. It is interesting that Harutyunyan and  Sedrakian \cite{Israeli} find (fig. 8) that even in the range, $B \sim 10^{12-14}$ G in the non quantizing regime, the above $B$ dependences hold approximately for the transverse conductivity: $ \sigma_{xx} = \sigma_{yy}   \sim 1/B^2$ and $\sigma_{xy} = -\sigma_{yx}$ goes as $1/B$ \cite{Israeli}. Furthermore, we also find from their figure that the diagonal transverse conductivity is roughly proportional to the square of the mass density whereas the non-diagonal transverse conductivity is proportional to the mass density.
\newline

\section{Discussion}

In any magnetar model the currents that determine the polar magnetic fields are in the plane transverse to the field.  We find that the diagonal transverse conductivity has a  $1/ B^2$ dependence. Such a strong dependence on the magnetic field would imply that magnetar crusts can have a conductivity that can be even $10^6$ times lower than those for pulsars. 

At pulsar valued magnetic fields ( $ \sim 10^{12} $ G) the outer electrons form a Fermi sea, as is the case for metals and the crust is a  crystal  in which the gravitational pressure is balanced by  the coulomb repulsion between the ions.  But when the magnetic fields go up ($ 10^{15(14)}$ G), the electrons in the outer shell of the crust (  for density $ \sim 10^7 $gm/cc) are localised, approximately in a landau radius, which is much smaller than the inter ionic spacing.  We  then have get deformed cigar like ions which are almost neutral and thus  deprived of adequate Coulomb repulsion. So we have a spaghetti like 1 D crystal along the field lines that is not tethered in the transverse plane. Since these localised electrons  cannot support transverse currents, the outer shell of the crust behaves like an transverse  insulator. Along  the field lines we have a regular 1 D conductor.

Therefore, if we have an initial state, with a high polar magnetic field, it will be supported by 
transverse currents that exist only below this thin shell at the surface. However, the resistivity will come down as we move to the interior of higher density . These currents will suffer ohmic dissipation and simultaneously the polar fields will gradually diminish. The heat so generated can  escape only  along the direction of the field lines via electron motion. Thus,  thermal transport is suppressed as it is highly anisotropic and one directional. Furthermore, the temperatures usually  associated with magnetar surfaces, $ < 10^7 $ K, are much lower than the Fermi energy which is of order , 1 Mev   $\sim 10^{10}$ K - thus there is also a Fermi suppression factor, $ k_T/E_f $.  This heat would then accumulate below the shell heating it up from below.
 As the below surface temperatures grows, thermal agitation from the heating layer below could set up  phonon like  modes in the spaghetti to dissipate the heat which could also  destabilize the spaghetti and give rise to bursting phenomena. Gradually, as the field goes down below , $10^{14}$ G, the spaghetti insulator will also  thin out and allow heat to exit.

In the dynamo model, we start with a high polar magnetic field and so we expect the ground state in the outer crust to be  a spaghetti insulator, where we may expect to see  the above phenomena.
On the other hand, in the screened core model, the field rises as it emerges out, cleaving the crystal and flaring. It is only at the later evolution that the crust becomes a mushy spaghetti insulator. In the screened charge model only when the polar field gets high does this effect step in, but by then most of the heat and screening currents have dissipated.

We have given a simple model of the ground state of matter in magnetar crusts and the magnetic field dependence of conductivity of the crust.
 It is also provides new and  interesting insights into the the properties of matter at very high magnetic fields particularly in the outer crust of magnetars. 
\newline 

\section{Acknowledgement}

We thank G. Baskaran, Dipankar Bhattacharya, Sameer Patel and Sajal Gupta for discussions.

\begin{thebibliography}{9}
	
	\bibitem{Hurley} Hurley, K, Nature, 434, 1098 (2005)
	
	\bibitem{Palmer} Palmer, D.M. et al, Nature, 434, 1107 (2005)
	
	\bibitem{Thompson} Thompson, C.,  Lyutikov, M. and  Kulkarni, S. R.,
	ApJ, 574, 332 (2002)
	
	\bibitem{DuncanThompson} Duncan, R. C. and Thompson, C., ApJ ,  392, L9 (1992)
	
	\bibitem{ThompsonDuncan} Thompson, C. and Duncan, R. C., MNRAS, 275, 255 (1995)	
	
	\bibitem{DarDR} Dar, A., \& De R\'ujula, A. 2000, in Results and Perspectives in Particle Physics (Ed. Mario Greco) Vol. 17, 13
	
	\bibitem{Dar} Dar, A., \textit{A\&A Suppl.} 138, 505 (1999)
	
	\bibitem{Marsden}   Marsden, D., Rothschild, R.E. and  Lingenfelter, R. E., ApJ, 520, L107 (1999)
	
	\bibitem{Kaspi}  Livingstone, M. A. et al,  The Astrophysical Journal, Volume 730, Number 2 (2011)
	
	\bibitem{DipSoni} Bhattacharya, D. and  Soni, V., arXiv:0705.0592v2 (2007)
	
	\bibitem{mag}  Soni V. and Haridass, N. D., MNRAS 425 (2) 1558-1566 (2012).
	
	\bibitem{Soni3}  Soni, V. et al,  MNRAS 482, (4) (2019)
	
	\bibitem{ShaRed} Sharma, R.and  Reddy, S., Phy. Rev. C , 83 025803 (2011)
	
	\bibitem{Bed}  Bedaqu,e P. F., et al  Phys. Rev. C, 88 (50, 055801) (2013) 
	
	
	\bibitem{shapiro} Shapiro, S. L. and Dong Lai, ApJ 383, 745, (1991)
	
	\bibitem{Dong Lai} Dong Lai, Reviews of Modern Physics, Vol 73  (2001)
	
	\bibitem{Baym}  Gordon Baym  NORDITA Lectures, Neutron Stars and the Properties of Matter at high density, Nordita and NBI (1977)
	
	\bibitem{Haensel}  Haensel, P. et al, A\&A, 229, 133, (1990) 
	
	
	
	\bibitem{Israeli}  Harutyunyan, A. and  Sedrakian, A.,
	Phys. Rev. C 94, 025805 (2016)
	
	
\end{thebibliography}

\bibliographystyle{abbrv}

\end{document}